\documentclass[%
 reprint,
 amsmath,amssymb,
 aps,
]{revtex4-2}

\usepackage{graphicx}
\usepackage{dcolumn}
\usepackage{bm}
\usepackage{hyperref}
\hypersetup{
	colorlinks=true,
	citecolor=blue,
	linkcolor=blue,
	urlcolor=blue
}
\usepackage{float}
\usepackage{balance}
\makeatletter
\renewcommand{\maketag@@@}[1]{\hbox{\m@th\normalsize\normalfont#1}}%
\makeatother
\usepackage{setspace}
\begin{document}
\title{A tale of two emergent games: opinion dynamics in dynamical directed networks}

\author{Yakun Wang}
\author{Bin Wu\thanks{*Corresponding author}}
\email{bin.wu@bupt.edu.en}
\affiliation{School of Science, Beijing University of Posts and Telecommunications, Beijing 100876, China.}

\begin{abstract}
Uni-directional social interactions are ubiquitous in real social networks whereas undirected interactions are intensively studied. We establish a voter model in a dynamical directed network. We analytically obtain the degree distribution of the evolving network at any given time. Furthermore, we find that the average degree is captured by an emergent game. On the other hand, we find that the fate of opinions is captured by another emergent game. Beyond expectation, the two emergent games are typically different due to the unidirectionality of the evolving networks. The Nash equilibrium analysis of the two games facilitates us to give the criterion under which the minority opinion with few disciples initially takes over the population eventually for in-group bias. Our work fosters the understanding of opinion dynamics ranging from methodology to research content.
\end{abstract}

\maketitle

\section{\label{sec:level1}Introduction}
Evolutionary dynamics on dynamical networks are essential to understand the complex systems in the real world including opinions, behaviours, and epidemics spreading \cite{RMP_statistical_physics, PRL_opinion_dynamics_1, PRL_opinion_dynamics_2, PRX_binary_state, PRL_2006_coevolution, PRL_2005_cooperation, Biosystems_a_mini_review, Gross_2007_review, Nature_evolutionary_dynamics, PRL_epidemic_dynamics, PLOS_2010_linking, JTB_2019_Kurokawa}. Typically, there are two sides of the coin for dynamics on dynamical networks: one is the fate of the evolutionary dynamics; the other is the ever-changing network. For opinion dynamics, the fate of opinions (consensus or polarization) has been investigated extensively \cite{Vazquez_PRL_2008, PRE_adaptive_gross_1, Emergence_polarization_PRL, PRL_Modeling}. In the empirical phenomena, the minority can win sometime, such as in the presidential elections \cite{polarization_PRX_Wang} and corporate operations \cite{corporate_operations}. What is the impact of social rewiring on the minority winning? The ever-changing network structure is much less investigated than the fate of opinions \cite{TAC_opinion_dynamics_1, CCC_2020_opinion_game, CPB_2022_voter_model}. It is still unclear whether one side of the coin can be inferred from the other. In particular, if the transient topology is captured, could it be used to predict the opinion formation? For opinion dynamics, say whether the opinion with more disciples initially can invade successfully in the end?

The voter model mirrors the simple average rule typically addressed in opinion dynamics in a stochastic manner \cite{PRL_voter_model_1, PRL_voter_model_2}. There is an amount of prior works studying the co-evolution of opinions on undirected networks \cite{Newman_2006_PRE, Daichi_2008_PRE, PNAS_2012_Richard}. However, unidirectional interactions are more ubiquitous in social relationships \cite{PNAS_directed, Twitter_directed, NC_2020_directed} and ecosystems \cite{Wolf_pack}. Compared with undirected networks, works on uni-directed networks are small in number \cite{PRE_gross_directed,  PNAS_2022_directed}. Simulations show that counter-intuitive results can arise due to the uni-directionality \cite{PRE_opinion_evolves_2}. On the other hand, in-group bias refers to the tendency for individuals to favor their own group over others' \cite{1971Social, 1979In, 1998intergroup, Rajiv_inbook, Opinion_Science}. The interaction between in-group bias and network structure can also lead to non-trivial dynamics, even for undirected networks \cite{Interface_in_group_bias}. Theoretical explanations are lacking if both the directionality of networks and in-group bias are taken into account. 

In this paper, we establish a voter model in a dynamical directed network. We address both sides of the coin, i.e., transient topology and the fate of opinions. We find that each side of the coin is captured by an \emph{emergent} game. In particular, the two games are typically different for dynamical uni-directional networks, whereas the two games are the same for dynamical undirected networks. This game perspective facilitates us to predict under what condition the minority with few disciples initially can win.

\section{Model}
Let us consider a system of $N$ individuals. The social relationships between individuals are captured by the dynamical directed networks. Each individual has on average $L$ incoming links and $L$ outgoing links. We assume that $N \gg L$. It implies that each individual has a limited number of neighbors compared with the population size. Each individual holds either opinion $+$ or opinion $-$. There are four types of links in the population, i.e., $S \buildrel \Delta \over = \left\{ {\overrightarrow { +  + } ,\overrightarrow { +  - } ,\overrightarrow { -  + } ,\overrightarrow { -  - } } \right\}$. For the directed link $\overrightarrow {XY}  \in S$, $X$ is the student and $Y$ is the teacher. Each individual has a student-node-set whose nodes flow into her and a teacher-node-set whose nodes flow out of her. We denote $x_{\scriptsize \pm}$ as the fraction of opinion $\pm$ in the population.

\begin{figure}[h]
	\includegraphics[scale=0.52]{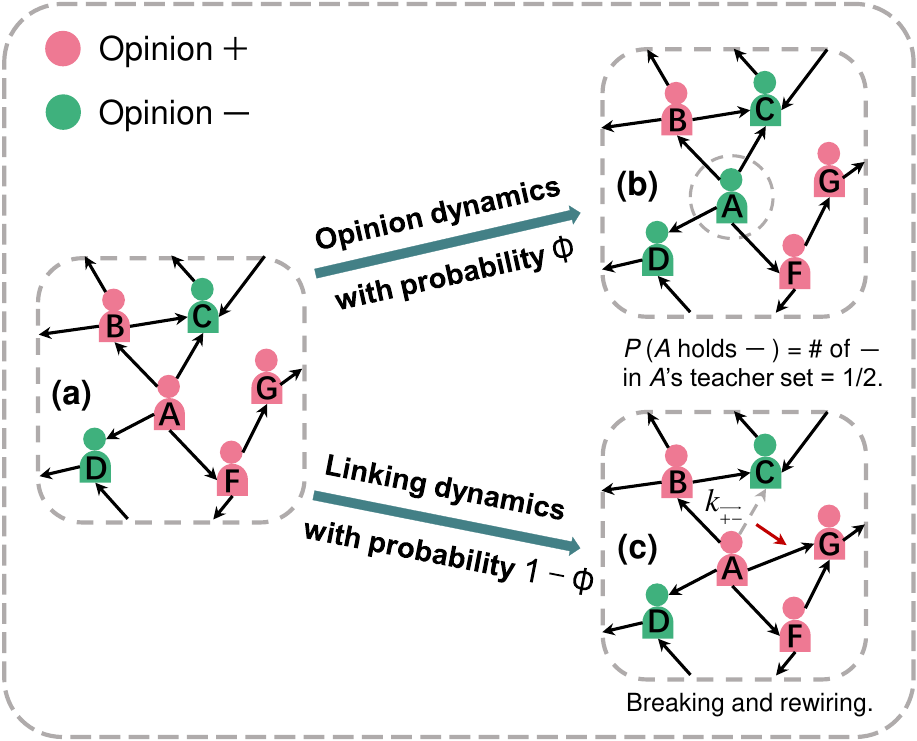}
	\caption{ \textbf{Coevolutionary dynamics of opinions and directed social relationships.} \textbf{(a)} The social relationships are described by the directed network. \textbf{(b)} With probability $\phi$, opinion updating happens. An individual is randomly selected to update her opinion. We assume $A$ is chosen. She learns from her teachers $B, C, D$ and $F$. Based on the voter model, $A$ adopts opinion $-$ with probability 1/2. \textbf{(c)} With probability $1-\phi$, linking dynamics happens. A directed link is selected randomly. We assume $\protect\overrightarrow {AC} $ is selected and $A$ is chosen. $A$ breaks the directed link with probability $k_{\protect\overrightarrow { +  - } }$ and rewires to $G$. }
	\label{linking_opinion}
\end{figure}

Opinion dynamics happens with probability $\phi $ and linking dynamics occurs with probability $1 - \phi $ at each time step \cite{CCC_2020_opinion_game, CPB_2022_voter_model, Shan_Social} (see Fig. \hyperref[linking_opinion]{1}). It is a coin tossing issue, in which opinion dynamics is the head whereas the linking dynamics is the tail. They are codependent. 

For \textbf{\itshape{opinion dynamics}}, we focus on the voter model \cite{PRX_binary_state}. An individual is randomly selected from the population. The probability that she adopts opinion $+$ is proportional to the number of teachers with opinion $+$ in her teacher-node-set. It is notable that if her teacher-node-set is empty, then she holds her current opinion. Opinion keeping is not a prior assumption on the characteristics of individuals in contrast with the zealots \cite{PRL_zealot, PRE_Zealotry}. 

For \textbf{\itshape{linking dynamics}}, there are three steps as follows:

\begin{itemize}
	\item[(i)] \emph{Selecting}. A directed link $\overrightarrow {XY} $ is randomly selected, where $\overrightarrow {XY}  \in S$. Either $X$ or $Y$ is randomly chosen. 
	\item[(ii)] \emph{Breaking}. The selected individual breaks the directed link $\overrightarrow {XY} $ with a pre-defined breaking probability ${k_{\scriptsize\overrightarrow {XY} }}$, where $0 < {k_{\scriptsize\overrightarrow {XY} }} < 1$. 
	\item[(iii)] \emph{Rewiring}. If the selected individual breaks the directed link $\overrightarrow {XY} $, then she rewires to a new individual, who is neither in her teacher-node-set nor in her student-node-set. 
\end{itemize}

\section{An emergent game for the transient topology}
For $\phi = 1$, the social relationships are invariant and individuals only update their opinions \cite{PRE_opinion_evolves_1, PRE_opinion_evolves_2}. For $\phi = 0$, the social network evolves all the time whereas the fractions of opinions are constant. We focus on $\phi \to 0 ^ + $. Individuals prefer to adjust their social relationships rather than to change their opinions, which is widespread in real social systems \cite{Network_evolves}. It leads to a time scale separation, that is, all the directed links are almost in the stationary regime when the opinion update occurs. The stationary distribution of the directed links is ${\pi_S} = \left( {{\pi _{\scriptsize\overrightarrow { +  + } }},{\pi _{\scriptsize\overrightarrow { +  - } }},{\pi _{\scriptsize\overrightarrow { -  + } }},{\pi _{\scriptsize\overrightarrow { -  - } }}} \right)$ (see \hyperref[Appendix A]{Appendix A} for details).

\begin{figure}
	\centering
	\includegraphics[scale=0.75]{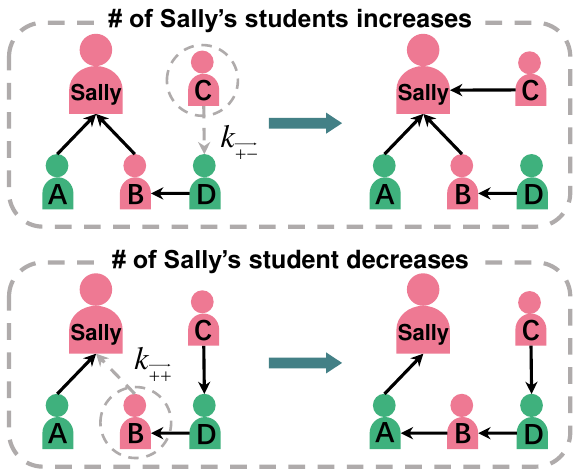}
	\caption{\textbf{Markov transitions of Sally's student size.} (Top Panel) Select one link which is not point to Sally. We assume that $\protect\overrightarrow {CD} $ is selected and $C$ is chosen to break the $\protect\overrightarrow {CD} $. Then $C$ chooses the new teacher Sally. Hence, the number of Sally's students increases by one. (Bottom Panel) Select one link which is point to Sally. We assume that the directed link $\protect\overrightarrow {BS} $ is selected and $B$ is chosen to break the $\protect\overrightarrow {BS} $. Then $B$ finds a new teacher $A$. Hence, the number of Sally's students decreases by one. }
	\label{markov_transitions_indegree}
\end{figure}

What are the key topology features that pave the way for successful invasions? We concentrate on the size of student-node-set, i.e., in-degree. Denote $d_ {{\rm{in}}\,\scriptsize +}$ as the in-degree for one node with opinion $+$. Suppose there is an individual, named after Sally. Without loss of generality, we assume that she adopts opinion $+$ and she has ${d_ {{\rm{in}}\,\scriptsize +} \in \left[ {0,N - 1} \right]}$ students. If an individual who is not Sally's current student rewires to Sally, then ${d_ {{\rm{in}}\,\scriptsize +} }$ increases by one with probability (see Fig. \hyperref[markov_transitions_indegree]{2})

\begin{equation}
\label{p_d_in+}
P_{{d_ {{\rm{in}}\,\scriptsize +} }}^ +  = \underbrace {\frac{{NL - {d_ {{\rm{in}}\,\scriptsize +} }}}{{NL}}}_{\,\scriptstyle{\rm{select}}\;{\rm{a}}\;{\rm{link}}\;{\rm{which}}\;\hfill\atop
	\scriptstyle{\rm{is}}\;{\rm{not}}\;{\rm{point}}\;{\rm{to}}\;{\rm{Sally}}\hfill}\underbrace {{\pi _S} \cdot \left( {\begin{array}{*{20}{c}}
		{{k_{\scriptsize\overrightarrow { +  + } }}/2}\\
		{{k_{\scriptsize\overrightarrow { +  - } }}/2}\\
		{{k_{\scriptsize\overrightarrow { -  + } }}/2}\\
		{{k_{\scriptsize\overrightarrow { -  - } }}/2}
		\end{array}} \right)}_{{\rm{break}}\;{\rm{the}}\;{\rm{link}}}\;\underbrace {\frac{1}{{N - 1}}}_{\scriptstyle{\rm{rewire}}\;{\rm{to}}\;{\rm{Sally}}\hfill}.
\end{equation}

\noindent On the other hand, if Sally's student rewires to other individuals, then ${d_ {{\rm{in}}\,\scriptsize +} }$ decreases by one with probability (see Fig. \hyperref[markov_transitions_indegree]{2})
\begin{equation}
\label{p_d_in-}
\!P_{{d_ {{\rm{in}}\,\scriptsize +} }}^ -  \!=\!\!\!\!\!\!\!\!\!\!\!\! \underbrace {\frac{{{d_ {{\rm{in}}\,\scriptsize +} }}}{{NL}}}_{\scriptstyle{\rm{select}}\;{\rm{a}}\;{\rm{link}}\;{\rm{which}}\;\hfill\atop
	\scriptstyle{\;\;\rm{is}}\;{\rm{point}}\;{\rm{to}}\;{\rm{Sally}}\hfill}\!\!\!\!\!\!\underbrace {\left( {\frac{{{\pi _{\scriptsize\overrightarrow { +  + } }}{k_{\scriptsize\overrightarrow { +  + } }}}/2}{{{\pi _{\scriptsize\overrightarrow { +  + } }} + {\pi _{\scriptsize\overrightarrow { -  + } }}}} + \frac{{{\pi _{\scriptsize\overrightarrow { -  + } }}{k_{\scriptsize\overrightarrow { -  + } }}}/2}{{{\pi _{\scriptsize\overrightarrow { +  + } }} + {\pi _{\scriptsize\overrightarrow { -  + } }}}}} \right)}_{{\rm{break}}\;{\rm{the}}\;{\rm{link}}}\!\underbrace 1_{\scriptstyle{\;\;\rm{rewire}}\;{\rm{to}}\hfill\atop
	\scriptstyle{\rm{other}}\;{\rm{nodes}}\hfill}.
\end{equation}\\ 
The one-step transition matrix ${P}$ of the Markov process is thus obtained. The Markov chain is aperiodic and irreducible, thus ergodic. Hence it has a unique stationary distribution ${\Xi _D} = \left( {{\xi _0},{\xi _1},{\xi _2}, \cdots {\xi _{N - 1}}} \right)$ which is determined by ${\Xi _D}P = {\Xi _D}$ \cite{Stochastic_processes}. Based on \cite{Stochastic_evolutionary_game_dynamics}, the stationary distribution is given by 
${\xi _j} = \left( {\frac{{P_0^ + }}{{P_j^ - }}\prod\nolimits_{i = 1}^{j - 1} {\frac{{P_i^ + }}{{P_i^ - }}} } \right){\left( {1 + \sum\nolimits_{k = 1}^{N - 1} {\frac{{P_0^ + }}{{P_k^ - }}\prod\nolimits_{i = 1}^{k - 1} {\frac{{P_i^ + }}{{P_i^ - }}} } } \right)^{ - 1}}$, where $1 \le j \le N - 1$ and $\prod\nolimits_{i = 1}^0 {\frac{{P_i^ + }}{{P_i^ - }}}  = 1$. For $j = 0$, we have ${\xi _0} = {\left( {1 + \sum\nolimits_{k = 1}^{N - 1} {\frac{{P_0^ + }}{{P_k^ - }}\prod\nolimits_{i = 1}^{k - 1} {\frac{{P_i^ + }}{{P_i^ - }}} } } \right)^{ - 1}}$. If the population size is infinitely large, i.e., $N \to + \infty $, then the in-degree follows the Poisson distribution (see more details in Supplemental Material). The obtained Poisson distribution facilitates us to obtain the average in-degree analytically. The expectation of ${d_ {{\rm{in}}\,\scriptsize +} }$ is $L{U_+}$, where $L$ is the average in-degree of the network and 
\begin{equation}
\label{average_indegree_U}
{U_ + } = {{{\pi _S} \cdot \left( {\begin{array}{*{20}{c}}
			{{k_{\scriptsize\overrightarrow { +  + } }}}\\
			{{k_{\scriptsize\overrightarrow { +  - } }}}\\
			{{k_{\scriptsize\overrightarrow { -  + } }}}\\
			{{k_{\scriptsize\overrightarrow { -  - } }}}
			\end{array}} \right)} \mathord{\left/
		{\vphantom {{{\pi _S} \cdot \left( {\begin{array}{*{20}{c}}
						{{k_{\scriptsize\overrightarrow { +  + } }}}\\
						{{k_{\scriptsize\overrightarrow { +  - } }}}\\
						{{k_{\scriptsize\overrightarrow { -  + } }}}\\
						{{k_{\scriptsize\overrightarrow { -  - } }}}
						\end{array}} \right)} {\left( {\frac{{{\pi _{\scriptsize\overrightarrow { +  + } }}{k_{\scriptsize\overrightarrow { +  + } }} + {\pi _{\scriptsize\overrightarrow { -  + } }}{k_{\scriptsize\overrightarrow { -  + } }}}}{{{\pi _{\scriptsize\overrightarrow { +  + } }} + {\pi _{\scriptsize\overrightarrow { -  + } }}}}} \right)}}} \right.
		\kern-\nulldelimiterspace} {\left( {\frac{{{\pi _{\scriptsize\overrightarrow { +  + } }}{k_{\scriptsize\overrightarrow { +  + } }} + {\pi _{\scriptsize\overrightarrow { -  + } }}{k_{\scriptsize\overrightarrow { -  + } }}}}{{{\pi _{\scriptsize\overrightarrow { +  + } }} + {\pi _{\scriptsize\overrightarrow { -  + } }}}}} \right)}}
\end{equation} 

\noindent which is the ratio of breaking the link in Eq. \hyperref[p_d_in+]{(1)} and Eq. \hyperref[p_d_in-]{(2)}. Interestingly, ${U_ {\scriptsize +} } = ({{{\pi _{\scriptsize\overrightarrow { +  + } }} + {\pi _{\scriptsize\overrightarrow { -  + } }}}})/x_+ = {{{f_ {\scriptsize +} }} \mathord{\left/
		{\vphantom {{{f_ {\scriptsize +} }} {\left( {{x_ {\scriptsize +} }{f_ {\scriptsize +} } + {x_ {\scriptsize -} }{f_ {\scriptsize -} }} \right)}}} \right.
		\kern-\nulldelimiterspace} {\left( {{x_ {\scriptsize +} }{f_ {\scriptsize +} } + {x_ {\scriptsize -} }{f_ {\scriptsize -} }} \right)}}$, where ${f_ {\scriptsize +} } = {{{x_ {\scriptsize +} }} \mathord{\left/
		{\vphantom {{{x_ {\scriptsize +} }} {{k_{\scriptsize\overrightarrow { +  + } }}}}} \right.
		\kern-\nulldelimiterspace} {{k_{\scriptsize\overrightarrow { +  + } }}}} + {{{x_ {\scriptsize -} }} \mathord{\left/
		{\vphantom {{{x_ {\scriptsize -} }} {{k_{\scriptsize\overrightarrow { -  + } }}}}} \right.
		\kern-\nulldelimiterspace} {{k_{\scriptsize\overrightarrow { -  + } }}}}$ and ${f_ {\scriptsize -} } = {{{x_ {\scriptsize +} }} \mathord{\left/
		{\vphantom {{{x_ {\scriptsize +} }} {{k_{\scriptsize\overrightarrow { +  - } }}}}} \right.
		\kern-\nulldelimiterspace} {{k_{\scriptsize\overrightarrow { +  - } }}}} + {{{x_ {\scriptsize -} }} \mathord{\left/
		{\vphantom {{{x_ {\scriptsize +} }} {{k_{\scriptsize\overrightarrow { -  - } }}}}} \right.
		\kern-\nulldelimiterspace} {{k_{\scriptsize\overrightarrow { -  - } }}}}$. ${f_ {\scriptsize \pm } }$ can be regarded as the payoff of the following game

\begin{equation}
{M_{{\rm{in \mbox{-} degree}}}} = \begin{array}{*{20}{c}}
{}&{\begin{array}{*{20}{c}}
	+ &{}&&{} -
	\end{array}}\\
{\begin{array}{*{20}{c}}
	+ \\
	{}\\
	-
	\end{array}}&{\left( {\begin{array}{*{20}{c}}
		{\displaystyle\frac{1}{{{k_{\scriptsize\overrightarrow { +  + } }}}}}&{\displaystyle\frac{1}{{{k_{\scriptsize\overrightarrow { -  + } }}}}}\\
		{\displaystyle\frac{1}{{{k_{\scriptsize\overrightarrow { +  - } }}}}}&{\displaystyle\frac{1}{{{k_{\scriptsize\overrightarrow { -  - } }}}}}
		\end{array}} \right)}
\end{array}.
\label{M_in_degree_matrix}
\end{equation}

\noindent The Nash equilibrium of Eq. \hyperref[M_in_degree_matrix]{(4)} $x_{\rm{in \mbox{-} degree} \kern 1pt \scriptsize + }^ * $ is

\begin{equation}
x_{\rm{in \mbox{-} degree} \kern 1pt \scriptsize + }^ * = \displaystyle\frac{{{1 \mathord{\left/
				{\vphantom {1 {{k_{\scriptsize\overrightarrow { -  - } }}}}} \right.
				\kern-\nulldelimiterspace} {{k_{\scriptsize\overrightarrow { -  - } }}}} - {1 \mathord{\left/
				{\vphantom {1 {{k_{\scriptsize\overrightarrow { -  + } }}}}} \right.
				\kern-\nulldelimiterspace} {{k_{\scriptsize\overrightarrow { -  + } }}}}}}{{{1 \mathord{\left/
				{\vphantom {1 {{k_{\scriptsize\overrightarrow { +  + } }}}}} \right.
				\kern-\nulldelimiterspace} {{k_{\scriptsize\overrightarrow { +  + } }}}} - {1 \mathord{\left/
				{\vphantom {1 {{k_{\scriptsize\overrightarrow { +  - } }}}}} \right.
				\kern-\nulldelimiterspace} {{k_{\scriptsize\overrightarrow { +  - } }}}} - {1 \mathord{\left/
				{\vphantom {1 {{k_{\scriptsize\overrightarrow { -  + } }}}}} \right.
				\kern-\nulldelimiterspace} {{k_{\scriptsize\overrightarrow { -  + } }}}} + {1 \mathord{\left/
				{\vphantom {1 {{k_{\scriptsize\overrightarrow { -  - } }}}}} \right.
				\kern-\nulldelimiterspace} {{k_{\scriptsize\overrightarrow { -  - } }}}}}}.
\end{equation}

\begin{figure}[h]
	\centering
	\includegraphics[scale=0.55]{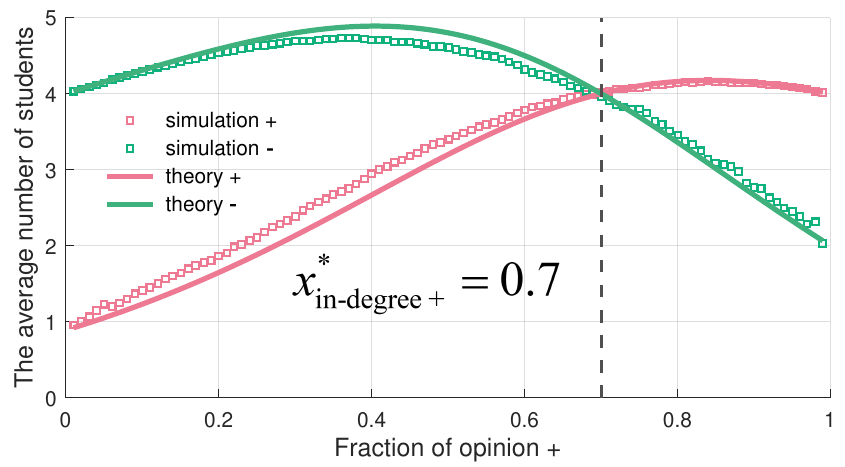}
	\caption{\textbf{Opinion $+$ has as many students as opinion $-$ does in the Nash equilibrium of ${M_{{\rm{in \mbox{-} degree}}}}$.} For in-group bias, if the proportion of opinion $+$ is larger than $ x_ {\rm in \mbox{-} degree \kern 1pt \scriptsize +} ^ * $, then the average degree of opinion $+$ is larger than opinion $-$'s. It implies that more students learn opinion $+$. Otherwise, the average degree of opinion $-$ is larger. Parameters: ${k _{\scriptsize\protect\overrightarrow { +  + } }} = 0.3$, ${k _{\scriptsize\protect\overrightarrow { +  - } }} = 0.6$, ${k _{\scriptsize\protect\overrightarrow { -  + } }} = 0.9$ and ${k _{\scriptsize\protect\overrightarrow { -  - } }} = 0.2$. $x_{\rm{in \mbox{-} degree}{\kern 1pt} \scriptsize + }^ * = 0.7$ in this case. We run 100 rounds of the simulation. We set $N = 100, L = 4$ and $\phi = 0.01$. }
	\label{average_indegree}
\end{figure}

\noindent It refers to a \emph{topology} in which opinion $+$ has as many students as opinion $-$ does. If ${x_ {\scriptsize +} } > x_ {\rm in \mbox{-} degree \kern 1pt \scriptsize +} ^ * $, the average in-degree of opinion $+$ is larger than that of opinion $-$ (see Fig. \hyperref[average_indegree]{3}). 

\section{Another emergent game for the fate of opinions}
For the evolutionary dynamics of opinions, the voter model in the evolving network is a Markov chain with state $x_{\scriptsize +}$ and the state space is $\left\{ {0,1/N,2/N, \cdots ,1} \right\}$ \cite{CCC_2020_opinion_game, CPB_2022_voter_model}. $x_{\scriptsize +}$ increases by $1 / N$ if an individual with opinion $-$ learns opinion $+$ with probability ${{q{\pi _{\scriptsize\overrightarrow { -  + } }}} \mathord{\left/
		{\vphantom {{q{\pi _{\scriptsize\overrightarrow { -  + } }}} {\left( {q{\pi _{\scriptsize\overrightarrow { -  + } }} + q{\pi _{\scriptsize\overrightarrow { -  - } }}} \right)}}} \right.
		\kern-\nulldelimiterspace} {\left( {q{\pi _{\scriptsize\overrightarrow { -  + } }} + q{\pi _{\scriptsize\overrightarrow { -  - } }}} \right)}} = {{{\pi _{\scriptsize\overrightarrow { -  + } }}} \mathord{\left/
		{\vphantom {{{\pi _{\scriptsize\overrightarrow { -  + } }}} {\left( {q{\pi _{\scriptsize\overrightarrow { -  + } }} + {\pi _{\scriptsize\overrightarrow { -  - } }}} \right)}}} \right.
		\kern-\nulldelimiterspace} {\left( {{\pi _{\scriptsize\overrightarrow { -  + } }} + {\pi _{\scriptsize\overrightarrow { -  - } }}} \right)}}$.
Here $q$ is the average size of the teacher set. Hence, the transition probability that $x_{\scriptsize +}$ increases by $1 / N$ is $T_{{x_ {\scriptsize +} }}^ +  = {x_ {\scriptsize -} }{\pi _{\scriptsize\overrightarrow { -  + } }}/\left( {{\pi _{\scriptsize\overrightarrow { -  + } }} + {\pi _{\scriptsize\overrightarrow { -  - } }}} \right)$. Similarly, the transition probability that $x {\scriptsize +}$ decreases by $1 / N$ is $T_{{x_ {\scriptsize +} }}^ -  = {x_ {\scriptsize +} }{\pi _{\scriptsize\overrightarrow { +  - } }}/\left( {{\pi _{\scriptsize\overrightarrow { +  + } }} + {\scriptsize\pi _{\overrightarrow { +  - } }}} \right)$. The probability that $x_ {\scriptsize +}$ remains constant is $T_{{x_{\scriptsize +}}}^0 = 1 - T_{{x_{\scriptsize +}}}^ +  - T_{{x_{\scriptsize +}}}^ - $. For large population size limit, i.e., $N \to +\infty $, the mean-field equation is given by ${\dot x_{\scriptsize +} } = T_{{x_{\scriptsize +}}}^ +  - T_{{x_{\scriptsize +}}}^ -$, capturing the evolution of the opinions. Thus we have ${\dot x_ {\scriptsize +} } = {\cal A} {x_ {\scriptsize +} }{x_ {\scriptsize -} }\left[ {\left( {{x_ {\scriptsize +} }{k_{\scriptsize\overrightarrow { +  - } }}/{k_{\scriptsize\overrightarrow { +  + } }} + {x_ {\scriptsize -} }} \right) - \left( {{x_ {\scriptsize +} } + {x_ {\scriptsize -} }{k_{\scriptsize\overrightarrow { -  + } }}/{k_{\scriptsize\overrightarrow { -  - } }}} \right)} \right]$, where ${\cal A} = {\left[ {{k_{\scriptsize\overrightarrow { +  + } }}{k_{\scriptsize\overrightarrow { -  - } }}\left( {{k_{\scriptsize\overrightarrow { +  - } }}{x_ {\scriptsize +} } + {k_{\scriptsize\overrightarrow { +  + } }}{x_ {\scriptsize -} }} \right)\left( {{k_{\scriptsize\overrightarrow { -  - } }}{x_ {\scriptsize +} } + {k_{\scriptsize\overrightarrow { -  + } }}{x_ {\scriptsize -} }} \right)} \right]^{ - 1}}$ is positive, provided that ${k_{\scriptsize\overrightarrow { +  + } }}{k_{\scriptsize\overrightarrow { +  - } }}{k_{\scriptsize\overrightarrow { -  + } }}{k_{\scriptsize\overrightarrow { -  - } }}{x_ {\scriptsize +} }{x_ {\scriptsize -} } \ne 0$. Multiplying ${\cal A} ^ { - 1 }$ on the right side does not alter the asymptotic dynamics, i.e., the fixed points and their stability. We end up with the equation
\begin{equation}
{\dot x_ {\scriptsize +} } = {x_ {\scriptsize +} }{x_ {\scriptsize -} }\left[ {\left( {{x_ {\scriptsize +} }{k_{\scriptsize\overrightarrow { +  - } }}/{k_{\scriptsize\overrightarrow { +  + } }} + {x_ {\scriptsize -} }} \right) - \left( {{x_ {\scriptsize +} } + {x_ {\scriptsize -} }{k_{\scriptsize\overrightarrow { -  + } }}/{k_{\scriptsize\overrightarrow { -  - } }}} \right)} \right],
\label{replicator_equation}
\end{equation}
which is a replicator equation with payoff matrix 
\begin{equation}
M_{\rm opinion } = \begin{array}{*{20}{c}}
{}&{\begin{array}{*{20}{c}}
	+ &{}&&{} -
	\end{array}}\\
{\begin{array}{*{20}{c}}
	+ \\
	{}\\
	-
	\end{array}}&{\left( {\begin{array}{*{20}{c}}
		{\displaystyle\frac{{k_{\scriptsize\overrightarrow { +  - } }}}{{{k_{\scriptsize\overrightarrow { +  + } }}}}}&1\\
		1&{\displaystyle\frac{{k_{\scriptsize\overrightarrow { -  + } }}}{{{k_{\scriptsize\overrightarrow { -  - } }}}}}
		\end{array}} \right)}
\end{array}.
\label{M_opinion_matrix}
\end{equation}
Intuitively, the payoff of an individual $+$ against an individual $+$ is proportional to ${k_{\scriptsize\overrightarrow { +  - }}}/{k_{\scriptsize\overrightarrow { +  + } }}$. If ${k_{\scriptsize\overrightarrow { +  - } }}$ increases, then the number of students with opinion $+$ who learn opinion $-$ decreases. A part of these students reconnect to new teachers with opinion $+$ and adopt opinion $+$. Hence the proportion of opinion $+$ increases. In our model, in-group bias corresponds to ${k _{\scriptsize\overrightarrow { +  - } }} > {k _{\scriptsize\overrightarrow { +  + } }}$ and $ {k _{\scriptsize\overrightarrow { -  + } }} > {k _{\scriptsize\overrightarrow { -  - } }}$. That is to say, students who adopt different opinions from their teachers' are more likely to change teachers than those who adopt the same opinions. The emergent payoff matrix in this case is a coordination game. There is only one unstable internal equilibrium for game Eq. \hyperref[M_opinion_matrix]{(7)} 
\begin{equation}
x_ {\rm opinion \kern 1pt \scriptsize +} ^ * = \frac{{{k_{\scriptsize\overrightarrow { -  + } }}/{k_{\scriptsize\overrightarrow { -  - } }} - 1}}{{{k_{\scriptsize\overrightarrow { +  - } }}/{k_{\scriptsize\overrightarrow { +  + } }} + {k_{\scriptsize\overrightarrow { -  + } }}/{k_{\scriptsize\overrightarrow { -  - } }} - 2}}.
\end{equation}
Thus all the individuals adopt opinion $+$ if the initial fraction of opinion $+$, denoted as $x_ {\rm initial \kern 1pt \scriptsize +}$, exceeds $x_ {\rm opinion \kern 1pt \scriptsize +} ^ *$. Otherwise all, the individuals reach consensus on opinion $-$. If $x_ {\rm initial \kern 1pt \scriptsize +}$ is between $1/2$ and $x_ {\rm opinion \kern 1pt \scriptsize +} ^ *$, opinion $-$ can win, even if opinion $-$ is minority initially, which is counter-intuitive. The emergent game helps to figure out when the minority can take over \cite{corporate_operations, polarization_PRX_Wang}.

Besides in-group bias, we also discuss other cases as follows [Fig. \hyperref[average_student]{4}].
\begin{itemize}
	\item \textbf{Out-group bias}: ${k _{\scriptsize\overrightarrow { +  - } }} < {k _{\scriptsize\overrightarrow { +  + } }}$ and $ {k _{\scriptsize\overrightarrow { -  + } }} < {k _{\scriptsize\overrightarrow { -  - } }}$. Eq. \hyperref[M_opinion_matrix]{(7)} refers to a coexistence game. $x_ {\rm opinion \kern 1pt \scriptsize +} ^ *$ is one internal stable equilibrium. Opinion $+$ and opinion $-$ coexist if they coexist in the beginning.
	\item \textbf{Dominance of opinion $+$}: ${k _{\scriptsize\overrightarrow { +  - } }} > {k _{\scriptsize\overrightarrow { +  + } }}$ and $ {k _{\scriptsize\overrightarrow { -  + } }} < {k _{\scriptsize\overrightarrow { -  - } }}$. Opinion $+$ is in-group bias, and the other opinion $-$ is out-group bias. $x^*=0$ is unstable and $x^*=1$ is stable. Then, opinion $+$ dominates the population. 
	\item \textbf{Dominance of opinion $-$}: ${k _{\scriptsize\overrightarrow { +  - } }} < {k _{\scriptsize\overrightarrow { +  + } }}$ and $ {k _{\scriptsize\overrightarrow { -  + } }} > {k _{\scriptsize\overrightarrow { -  - } }}$. Opinion $+$ is out-group bias, and the other opinion $-$ is in-group bias. $x^*=0$ is stable and $x^*=1$ is unstable. Then, opinion $-$ dominates the population. 
\end{itemize}

\section{A tale of two games to approach the counter-intuitive phenomenon}
If ${k_{\scriptsize\overrightarrow { +  - } }} = {k_{\scriptsize\overrightarrow { -  + } }} = k$, where $0 < k < 1$, we have $M_{\rm{opinion}} = k \cdot M_{\rm{in \mbox{-} degree}}$ and $x_{\rm{in \mbox{-} degree} \kern 1pt \scriptsize + }^ * = x_ {\rm opinion \kern 1pt \scriptsize +} ^ *$, which implies that \emph{ONE} emergent game is sufficient to capture both the fate of opinions and the transient topology. It mirrors an undirected-like network. The network has symmetric-like properties in a statistical sense although it is still directed. In this case, if one opinion has more students than the other initially, then the former opinion can take over the population eventually.

\begin{figure}[h]
	\centering
	\includegraphics[scale=0.27]{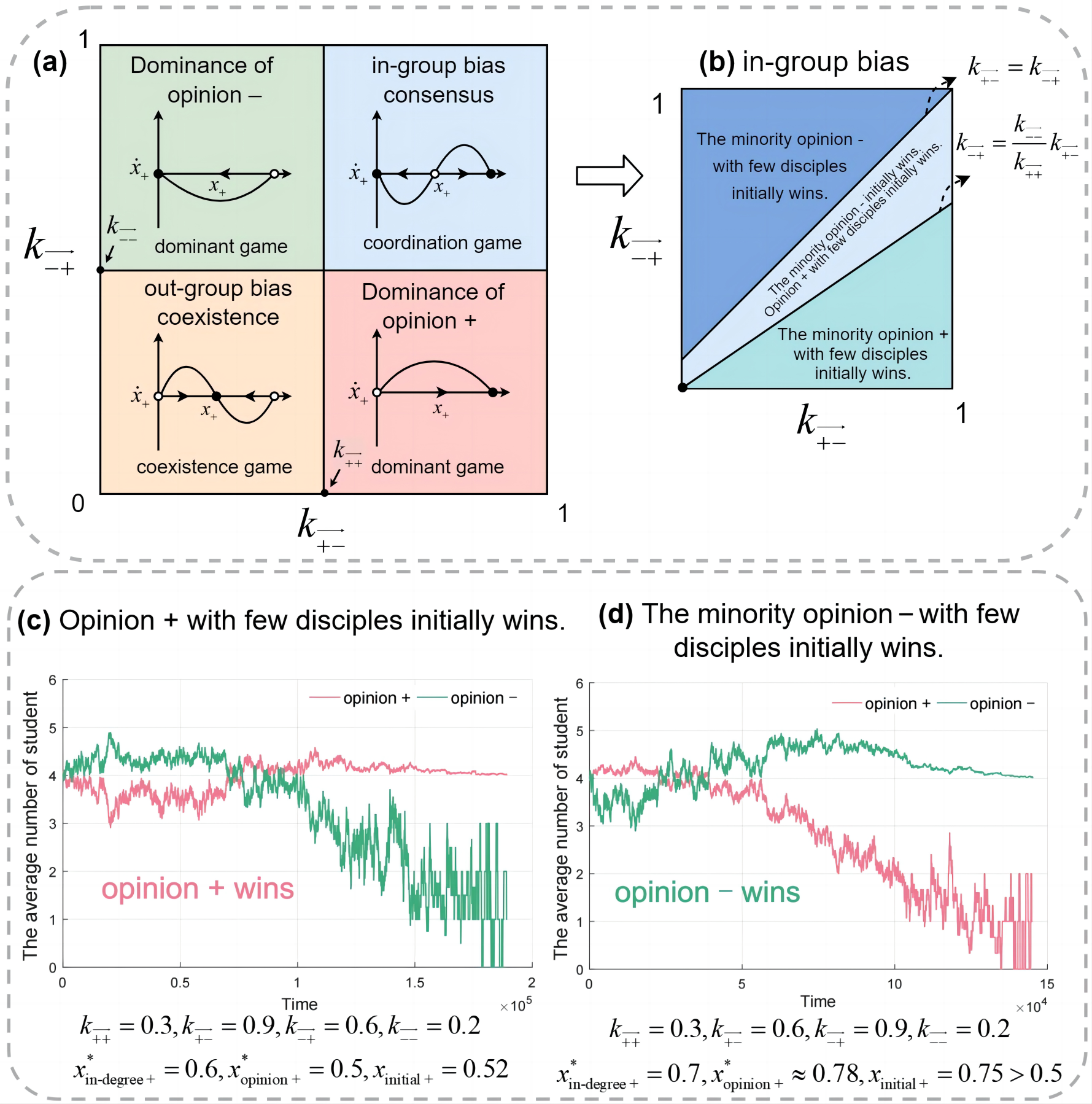}
	\caption{\textbf{A tale of two games.} \textbf{(a)} Game interactions for the fate of opinions. \textbf{(b)} Based on the two games, counter-intuitive phenomena emerge. \textbf{(c)} Opinion $+$ with few disciples in the beginning wins. Parameters: ${k _{\scriptsize\protect\overrightarrow { +  + } }} = 0.3$, ${k _{\scriptsize\protect\overrightarrow { +  - } }} = 0.9$, ${k _{\scriptsize\protect\overrightarrow { -  + } }} = 0.6$ and ${k _{\scriptsize\protect\overrightarrow { -  - } }} = 0.2$. $x_{\rm{in \mbox{-} degree}{\kern 1pt} \scriptsize + }^ * = 0.6$ and $x_{\rm{opinion}{\kern 1pt} \scriptsize + }^ * = 0.5$. The initial fraction of opinion $+$ is 0.52. \textbf{(d)} The minority opinion $-$ with few disciples in the beginning wins, which is even more counter-intuitive. Parameters: ${k _{\scriptsize\protect\overrightarrow { +  + } }} = 0.3$, ${k _{\scriptsize\protect\overrightarrow { +  - } }} = 0.6$, ${k _{\scriptsize\protect\overrightarrow { -  + } }} = 0.9$ and ${k _{\scriptsize\protect\overrightarrow { -  - } }} = 0.2$. $x_{\rm{in \mbox{-} degree}{\kern 1pt} \scriptsize + }^ * = 0.7$ and $x_{\rm{opinion}{\kern 1pt} \scriptsize + }^ * \approx 0.78$. The initial fraction of opinion $+$ is 0.75. We set $N = 100, L = 4$ and $\phi = 0.01$. }
	\label{average_student}
\end{figure}

If ${k_{\scriptsize\overrightarrow { +  - } }} \ne {k_{\scriptsize\overrightarrow { -  + } }}$, the emergent game $M_{\rm{in \mbox{-} degree}}$ differs from $M_{\rm{opinion}}$, which can give rise to some even more counter-intuitive results [Fig. \hyperref[average_student]{4(b)}]. For in-group bias and ${k_{\scriptsize\overrightarrow { +  - } }} > {k_{\scriptsize\overrightarrow { -  + } }}$, if $x_ {\rm initial \kern 1pt \scriptsize +} \in [x_ {\rm opinion \kern 1pt \scriptsize +} ^ *, x_ {\rm in \mbox{-} degree \kern 1pt \scriptsize +} ^ *]$, then opinion $+$ invades successfully in the end, even if opinion $+$ has few disciples [Fig. \hyperref[average_student]{4(c)}]. Hence, the number of disciples of an opinion is not the key factor for the successful invasion for the dynamical directed networks. Similarly, for in-group bias and ${k_{\scriptsize\overrightarrow { +  - } }} < {k_{\scriptsize\overrightarrow { -  + } }}$, if $x_ {\rm initial \kern 1pt \scriptsize +} \in [x_ {\rm in \mbox{-} degree \kern 1pt \scriptsize +} ^ *, x_ {\rm opinion \kern 1pt \scriptsize +} ^ *]$, then opinion $-$ invades successfully in the end, even if opinion $-$ has few disciples. Furthermore, if $x_ {\rm initial \kern 1pt \scriptsize +} > 1/2$ and $x_ {\rm initial \kern 1pt \scriptsize +} \in [x_ {\rm in \mbox{-} degree \kern 1pt \scriptsize +} ^ *, x_ {\rm opinion \kern 1pt \scriptsize +} ^ *]$, then opinion $-$ can take over in the end, even if the minority opinion $-$ with few disciples initially, which is even more counter-intuitive [Fig. \hyperref[average_student]{4(d)}]. 

\section{Conclusion and Discussion}
We focus on the voter model in the evolving directed network. The transient topology in the sense of average degree and the fate of opinions are found to be captured by two emergent 2$\times 2$ games, respectively. Therefore, we show that opinion dynamics is equivalent to games, both network wise and opinion wise. With a game perspective, our work not only provides the threshold under which the minority can win, but also gives the critical social adjustment rule to ensure that the minority with few disciples initially can win, which is even more counter-intuitive. This also implies that transient topology alone is not sufficient to predict the fate of opinions, unless the network is undirected.

Opinion dynamics and evolutionary game theory are two fields in complex systems. Recent years have seen an increasing interest in studying opinion dynamics via the game approach \cite{EPL_2016_opinion_game, TAC_opinion_game, Chaos_2022_opinion_game, RSOS_opinion_game}. All of the previous works assume a utility function (game interaction) first, and then individuals adjust their opinions via maximizing their payoffs. Individuals in our model have no payoff in mind when updating opinions. The game itself is a \emph{result}, rather than an assumption. Our work sheds a deeper connection between opinion dynamics and evolutionary games. Furthermore, we are the first to use an emergent game (not games assumed priorly) to capture the time-dependent average degree in contract with previous works \cite{Degree_01, Degree_02, Degree_03}. Our work thus bridges the gap between game theory and transient topology in dynamical networks. 

It is shown that if the network is undirected, then Eq. \hyperref[M_in_degree_matrix]{(4)} is equivalent to Eq. \hyperref[M_opinion_matrix]{(7)}. Consequently, \emph{ONE} emergent game is sufficient to capture both the transient topology and fate of opinions. It implies that the transient topology (student size) suffices to predict the opinion formation, that is, the opinion with more disciples wins eventually in the dynamical undirected networks. Hence, the unidirectionality of networks is a necessary condition to make opinions with few disciples take over.

Fragmentation does not play a role in our model unlike other opinion dynamics models \cite{Vazquez_PRL_2008, PRE_adaptive_gross_1, PRE_gross_directed}. We show that there are only two absorbing states for the approximated Markov chain. It implies that individuals reach consensus sooner or later. Furthermore, we study the complexity of the revised model, in which the previous step $(i)$ is replaced by the following: the two extremes of the selected link are chosen by different probabilities depending on the link type rather than randomly irrespective of the link type. We find that the transient topology in the revised model is fully captured by an emergent three-player two-strategy game which has at most two internal equilibria. On the other hand, the fate of opinions is captured by a replicator equation of an emergent four-player two-strategy game which has at most three internal equilibria. These multi-equilibria of the two emergent games imply that counter-intuitive results are more likely to take place than the previous model (see Supplemental Material). This highlights the complexity of the rewiring process on unidirectional networks is key for such the counter-intuitive opinion formation process. 

To sum up, based on the Nash equilibrium analysis of the two
emergent games, our work fills the gap between the literature on opinion dynamics and the counter-intuitive phenomenon.

\section*{acknowledgments}
We gratefully acknowledge Xunlong Wang, who inspires us to find that the in-degree follows the Poisson distribution in the infinite large population size limit. This work is supported by National Natural Science Foundation of China (NSFC) under Grant 61751301.

\section*{Appendix A: Linking Dynamics} \label{Appendix A}
Here the number of directed links $NL$ is constant. Each directed link
$i \left( {i = 1,2, \cdots ,NL} \right)$ is selected with probability $1/NL$. In time $t$, we randomly select a directed link ${i^t} = i$. If the selected ${i^t}$ does not break, then we have ${i^{t + 1}} = {i^t}$. Otherwise, a new directed link is introduced, denoted as ${i^{t + 1}}$. We denote the type of directed edge of ${i^t}$ by $T\left( {{i^t}} \right)$, where
$T\left( {{i^t}} \right) \in S$.

The linking dynamics is captured by Markov chain with transition matrix ${Q_{ ( {\scriptsize\overrightarrow {AB} } ) ( {\scriptsize\overrightarrow {CD} } )}}$, which is the probability that link $\overrightarrow {AB} $ transforms to link $\overrightarrow {CD} $ in one time step. For instance, ${Q_{\scriptsize ( {\overrightarrow { +  + } } ) ( {\overrightarrow { +  + } } ) }}$ is the probability that ${i^t}$ of type $\overrightarrow { +  + } $ transforms to ${i^{t + 1}}$ of type $\overrightarrow { +  + } $. In this case, one of the following two cases occurs:\\
(1) ${i^t}$ is not selected (with probability $\left( {NL - 1} \right)/NL$).\\
(2) ${i^t}$ is selected (with probability $1/NL$). Either the original $\overrightarrow { +  + } $ link is not broken (with probability $1 - {k_{\scriptsize\overrightarrow { +  + } }}$), or the original $\overrightarrow { +  + } $ link is broken and student $+$ (or teacher $+$) rewires to a new teacher(student) $+$ (with probability ${k_{\scriptsize\overrightarrow { +  + } }} x_{\scriptsize +}$, where $x_{\scriptsize +}$ is the fraction of opinion $+$). Hence,
\begin{equation}
\setcounter{equation}{1}
\renewcommand\theequation{A\arabic{equation}}
{Q_{\left( {\scriptsize\overrightarrow { +  + } } \right)\left( {\scriptsize\overrightarrow { +  + } } \right)}} = \frac{{NL - 1}}{{NL}} + \frac{1}{{NL}}\left( {1 - {k_{\scriptsize\overrightarrow { +  + } }} + {k_{\scriptsize\overrightarrow { +  + } }}{x_ {\scriptsize +} }} \right).
\end{equation}
The transition probability matrix is given by
\begin{equation}
\setcounter{equation}{2}
\renewcommand\theequation{A\arabic{equation}}
Q = \frac{{NL - 1}}{{NL}}{I_4} + \frac{1}{{NL}}V,
\end{equation}
where ${I_4}$ is the identity matrix and the matrix $V$ is given by 
\begin{equation}
\setcounter{equation}{3}
\renewcommand\theequation{A\arabic{equation}}
\begin{array}{l}
V = \\
\begin{array}{*{20}{c}}
{}&{\begin{array}{*{20}{c}}
	{\overrightarrow { +  + } }&{}&{}&\qquad{\overrightarrow { +  - } }&{}&{}&{}&\;\;\;{\overrightarrow { -  + } }&{}&{}&{}&{}&\;\;\;{\overrightarrow { -  - } }
	\end{array}}\\
{\begin{array}{*{20}{c}}
	{\overrightarrow { +  + } }\\
	{\overrightarrow { +  - } }\\
	{\overrightarrow { -  + } }\\
	{\overrightarrow { -  - } }
	\end{array}}&{\left( {\begin{array}{*{20}{c}}
		{1 - {k_{\scriptsize\overrightarrow { +  + } }}{x_ {\scriptsize -} }}&{{k_{\scriptsize\overrightarrow { +  + } }}{x_ {\scriptsize -} }/2}&{{k_{\scriptsize\overrightarrow { +  + } }}{x_ {\scriptsize-} }/2}&0\\
		{{k_{\scriptsize\overrightarrow { +  - } }}{x_ {\scriptsize +} }/2}&{1 - {k_{\scriptsize\overrightarrow { +  - } }}/2}&0&{{k_{\scriptsize\overrightarrow { +  - } }}{x_ {\scriptsize -} }/2}\\
		{{k_{\scriptsize\overrightarrow { -  + } }}{x_ {\scriptsize +} }/2}&0&{1 - {k_{\scriptsize\overrightarrow { -  + } }}/2}&{{k_{\scriptsize\overrightarrow { -  + } }}{x_ {\scriptsize -} }/2}\\
		0&{{k_{\scriptsize\overrightarrow { -  - } }}{x_ {\scriptsize +} }/2}&{{k_{\scriptsize\overrightarrow { -  - } }}{x_ {\scriptsize +} }/2}&{1 - {k_{\scriptsize\overrightarrow { -  - } }}{x_ {\scriptsize +} }}
		\end{array}} \right)}
\end{array},
\end{array}
\label{pingwenfenbu}
\end{equation}
The matrix $V$ is an approximation because it is possible that an individual reconnects to her students or teachers. Since the population size is much larger than the average degree of the nodes, i.e., $N \gg L$, the approximation is completely acceptable. The state space of the Markov chain is $S$. If $\phi \ll 1$ and ${k_{\scriptsize\overrightarrow { +  + } }}{k_{\scriptsize\overrightarrow { +  - } }}{k_{\scriptsize\overrightarrow { -  + } }}{k_{\scriptsize\overrightarrow { -  - } }}{x_ {\scriptsize +} }{x_ {\scriptsize -} } \ne 0$, there is a unique stationary distribution ${\pi_S} = \left( {{\pi _{\scriptsize\overrightarrow { +  + } }},{\pi _{\scriptsize\overrightarrow { +  - } }},{\pi _{\scriptsize\overrightarrow { -  + } }},{\pi _{\scriptsize\overrightarrow { -  - } }}} \right)$, where 

\begin{equation}
\setcounter{equation}{4}
\renewcommand\theequation{A\arabic{equation}}
\label{eq.A4}
{\pi _S} = {\cal N ^ * }\left( {\frac{{x_ {\scriptsize +} ^2}}{{{k_{\scriptsize\overrightarrow { +  + } }}}},\frac{{{x_ {\scriptsize +} }{x_ {\scriptsize -} }}}{{{k_{\scriptsize\overrightarrow { +  - } }}}},\frac{{{x_ {\scriptsize +} }{x_ {\scriptsize -} }}}{{{k_{\scriptsize\overrightarrow { -  + } }}}},\frac{{x_ {\scriptsize -} ^2}}{{{k_{\scriptsize\overrightarrow { -  - } }}}}} \right)
\end{equation}
\noindent
determined by equation ${\pi_S}Q = {\pi_S}$. ${\cal N ^ * } > 0$ is a normalization factor. Here ${\pi _{\overrightarrow {XY} }}$ is the probability of the directed link $\overrightarrow {XY} $ in the stationary regime. If $x_+$ increases, then $\pi_{\overrightarrow {+ +}}$ increases and $\pi_{\overrightarrow {- -}}$ decreases. That is, the number of $\overrightarrow {+ +}$ increases and the number of directed links $\overrightarrow {- -}$ decreases.
\\

\bibliography{PRE_two_games_xiugaiban}
\end{document}